\documentstyle[preprint,eqsecnum,aps,epsf]{revtex}
\tightenlines
\newcommand{\be}{\begin{equation}}
\newcommand{\ee}{\end{equation}}
\newcommand{\ba}{\begin{eqnarray}}
\newcommand{\ea}{\end{eqnarray}}
\begin{document}
\title{\Large{\bf{Inelastic Dissipation in Wobbling Asteroids and
Comets}}}
\author{\Large{Michael Efroimsky}}
\address{Department of Physics, Harvard University} 
\address{(NEW ADDRESS: Inst. for Mathematics, University of Minnesota)} 
\address{e-mail: efroimsk@ima.umn.edu}
\author{\Large{and}}
\author{\Large{A. Lazarian}}
\address{Department of Astronomy, University of Wisconsin, Madison WI 53706}
\address{e-mail: lazarian@dante.astro.wisc.edu}
\author{PUBLISHED IN:} 
\author{\large{\bf{{{Monthly Notes of the Royal Astronomical Society
(MNRAS)}}, ~Vol. {\bf{311}}, ~p. 269 (2000)}}}
\author{~\\}
\maketitle
\begin{abstract}
Asteroids and comets dissipate energy when they rotate about any axis
different from the axis of the maximal moment of inertia. 
We show that the most efficient internal relaxation happens
at twice the frequency of the body's precession. Therefore earlier
estimates that ignore the double frequency input underestimate the
internal relaxation in asteroids and comets. We suggest that the
Earth's seismological data may poorly represent the acoustic 
properties of asteroids and comets as internal relaxation increases
in the presence of moisture. At the same time, owing to the non-linearity
of inelastic relaxation, small angle nutations can persist for very long
time spans, but our ability to detect such precessions is limited by the
resolution of the radar-generated images.
Wobbling may provide valuable information on the
composition and structure of asteroids and on their recent history
of external impacts. 
\end{abstract}

\pagebreak

\section{Introduction}

Comet P/Halley exhibits a very complex rotational motion (e.g., Peale and 
Lissauer 1989), which is attributed to its rotation about an axis that
does not coincide with the axis of the major inertia. Some asteroids 
also wobble like, for example, 4179 Toutatis  
(Ostro et al. 1993, Harris 1994, Ostro et al. 1995, Hudson and Ostro 
1995, Scheeres et al. 1998, Ostro et al. 1999). 

A precessing body, as opposed to one steadily rotating about its axis
of maximal inertia, is subjected to alternating stresses. These
stresses deform the body, and cause energy dissipation due to inelastic
effects. Naturally, this cannot change the angular momentum of a
freely rotating body. Therefore the body tends to relax to the state
of minimum energy, namely to the state of rotation with the maximum-inertia 
axis parallel to the angular momentum vector.
External impacts may, however, excite wobble: if the body is observed 
precessing, this means that it has been subjected to impacts within the 
characteristic time of inelastic relaxation. Tidal interaction may be an 
additional source of wobble excitation. This source may be of a special 
importance for planet-crossing asteroids and comets (Black et al 1999). 
Jetting may be 
another reason for cometary precession. Generally, spinning bodies may start wobbling when they change their inertial axes through the loss of material (craters or outgassing). It is the balance between the 
impacts and tidal interactions, on the one hand, and the inelastic 
relaxation, on the other hand, that determines the dynamics of the body's  
rotation.

Therefore a study of the rotation of asteroids and comets may provide valuable
information about their recent history of external impacts (and tidal 
interactions), provided that we quantitatively understand the 
inelastic-dissipation process.

An important study of inelastic relaxation in a wobbling asteroid
has been performed by Burns \& Safronov (1973). In their 
treatment they decomposed the complex pattern of body
deformations into bending and bulge; calculated deformations and
the elastic energy associated with the deformations, and calculated
the energy-dissipation rate, using the material quality factor (so-called
$\; Q \;$ factor). 

Our results differ from those in
Burns and Safronov (1973) because, first, the dissipation at the 
double-frequency was missed in the latter study (which in fact provides the 
leading inpunt in the dissipation process)\footnote{Another attempt of 
calculating the stresses 
emerging in a precessing body, presented in a long-standing article (Purcell 
1979), also omitted the double-frequency contribution to the stress tensor. }. 
Second, we shall try to be more 
exact in our quest for the probable values of the quality factor $\; Q \;$. 
Burns and Safronov in their article quoted the standard seismological data 
($\; 10^2 < \; Q \; < \; 10^3 \;$), and eventually used $\; Q \; \approx \;
300 \;$ as a more-or-less-acceptable average. Their listed range of 
values of $\; Q \;$ is indeed typical for almost all terrestial rocks under
the conditions natural for the lithosphere: at temperatures varying 
from the room temperature up to several hundred Celsius; at pressures about 
several $\; MPa \;$; at frequencies from $\; 10^{-1} \; Hz \;$ through several
$\; kHz \;$; and most important, for rocks with traces of moisture. None of 
these conditions are believed to hold for tumbling asteroids and comets, and 
therefore we shall not be able to borrow data from the geophysical literature.

Uncertainties in the approach presented in the previous literature motivate 
our search for more rigorous methods, as well as for more appropriate values 
for the quality factor in order to provide more accurate estimates for the 
inelastic-relaxation time. 

This is the second paper in the series. In the first (Lazarian \& 
Efroimsky 1999) we calculated the internal relaxation in a precessing dust 
grain. In what follows we shall remind the reader basic facts about 
solid 
body rotation (section~2). We shall calculate the stresses caused by the 
precession, and the rates of internal relaxation (all these technical 
calculations are in the Appendix). Then we shall compare our expression for the
relaxation rate with that suggested by 
earlier researchers (section~3). Section~4 discusses whether our results are 
applicable not only to oblate but also to prolate bodies. A  
discussion of acoustical properties of materials is given in 
section~5. A particular example of wobble (asteroid 4179 Toutatis) is 
discussed in section~6, and the conclusions are summarised in section~7.

\section{Solid-Body Rotation}

We are interested in the free rotation of an oblate symmetric body. Its 
principal moments of inertia will be denoted by $I_i$. One may assume, without 
loss of generality, that 
\be
I_{3} \; > \; I_{1} \; = \; I_{2} \; \equiv \; I \; \; , 
\label{2.6}
\ee
The body's inertial angular velocity will be denoted by $\bf{\Omega}$, and its precession 
rate will be called $\bf{\omega}$.
The body-frame-related coordinate system is naturally associated with the 
three principal axes of inertia: $1$, $2$, and $3$, with coordinates denoted 
as $x$, $y$, $z$, and unit vectors ${\mathbf{e}}_{1}$, ${\mathbf{e}}_{2}$, 
${\mathbf{e}}_{3}$. The body-frame-related components of 
$\bf{\Omega}$ will be denoted as $\; \Omega_{1,2,3} \;$. 

The other, inertial, frame ($X$, $Y$, $Z$), with unit 
vectors ${\mathbf{e}}_{X}$, ${\mathbf{e}}_{Y}$, ${\mathbf{e}}_{Z}$, is chosen 
so that its $Z$ axis is parallel to the (conserved) angular momentum 
$\mathbf{J}$, and its origin coincides with that of the body frame (i.e., is 
at the body's centre of mass). Coordinates with respect to the inertial 
frame are denoted by the same capital letters as its axes: $X$, $Y$, and $Z$. 

We shall be interested on $\dot{\theta}$, the rate of the 
maximum-inertia axis' approach to the direction of angular momentum $\bf{J}$. 
 To achieve this goal, one has to know the rate of energy losses caused 
by the inelastic deformation. To calculate the deformation, 
we have to know 
the acceleration experienced by a particle located inside the body at a 
point ($x$, $y$, $z$). Note that we address the inertial acceleration, i.e.,
that with respect to the inertial frame $(X,Y,Z)$, but we express it in 
terms of coordinates $x$, $y$ and $z$ of the body frame $(1,2,3)$. 

The fast processes (revolution and precession of a symmetric oblate body) are 
described by the Euler equations whose solution, ignoring any slow 
relaxation, reads (Fowles and Cassiday 1986, Section 9.5; Landau and Lifshitz 1976):
\be
{\Omega}_1 \; \; = \; \; {\Omega}_{\perp} \cos {\omega}t~~,~~~
{\Omega}_2 \; \; = \; \; {\Omega}_{\perp} \sin {\omega}t~~,~~~
{\Omega}_3 \; \; = \; \; const
\label{rotation}
\ee
where
\be
{{\Omega}_{\perp}} \; \; \equiv \; \; {\Omega} \; \; \sin \; {\alpha}~~, ~~~~~
{{\Omega}_{3}} \; \; \equiv \; \; {\Omega} \; \; \cos \; {\alpha}
\label{24}  
\ee
and
\be
{{\Omega}_{\perp}}/{{\Omega}_{3}} \; \; = \; \; \tan \; \alpha \; \; = \; \; 
h \; \; \tan \; {\theta}~~~.
\label{25}
\ee
Here $\,\theta\,$ and $\;\alpha\;$ are the angles made by the maximal-inertia 
axis with $\bf{J}$ and $\,\bf{\Omega}\,$. The angular velocity ${\bf{\Omega}}$ nutates around the 
principal axis $3$ at a constant angular velocity 
\be
\omega \; = \; (h - 1) \Omega_3, 
\; \; \; \; \; \; \; \;  h\equiv {I_3}/I~~~. 
\label{9}
\ee
However from the point of view of the inertial observer it is rather 
axis $3$ that wobbles about $\bf{J}$. Thus the angle $\theta$ between  $\bf{J}$
and axis 3 is constant\footnote{The 
rate $\; \omega \;$ of 
precession is of the order of $|{\bf{\Omega}}|$, except in the case of 
$\;h\,\rightarrow\,1\;$, or in a very special case of 
${\bf{\Omega}}$ and ${\bf{J}}$ being orthogonal or almost orthogonal to the 
maximal-inertia axis $3$. Hence one may call the rotation and precession 
``fast motions'', implying that the relaxation is slow: $\; 
\dot{\theta}\ll\omega$. It is in this sense that we assume $\,\theta\,$ is  
constant}, as long as the energy is conserved.

The components of $\bf{\Omega}$ are connected with the absolute value of the 
angular momentum:
\be
{\Omega}_3 \; \; = \; \; \frac{J_3}{I_3} \; \; = \; \; \frac{J}{I_3} \; \; 
\cos \; \theta~~, ~~~~
{\Omega}_{\perp} \; \; = \; \; \frac{J}{I_3} \; h \; \; \sin \; \theta\;\;\;\;.
\label{om4}
\ee

As $\; \bf{\Omega} \;$ preceses with rate $\; \omega \;$, its 
components bear a time-dependence in the forms of $\; \sin \omega t \;$ and 
$\; \cos \omega t \;$. Since the centripetal component of the acceleration is 
quadratic in $\; \bf{\Omega} \;$, a double frequency must unavoidably emerge 
in the expression for acceleration.  
Unfortunately, this circumstance has been missed in the literature hitherto, 
and all the authors have been considering dissipation at the principal 
frequency $\; \omega \;$ solely. In our recent article (Lazarian and Efroimsky
1999) we presented a comprehensive calculation of the acceleration. The 
time-dependent and time-independent components of the acceleration give birth 
to time-dependent and time-independent components of the stresses, 
correspondingly. The stresses and strains arising at the double frequency 
considerably increase the elastic energy associated with the vibrations. This 
leads to a higher rate of energy dissipation in the body, and therefore to a 
much higher rate of relaxation. Calculation of the alignment rate, for an 
oblate body modelled by a prism of sizes $2a \times 2a \times 2c$, ($c < a)$, 
is presented in the Appendix. Here follows the final result:
\be
d \theta/dt \; = \; - \; \frac{3}{2^4} \; \sin^3 \theta \;\;\left[\;\frac{63\; 
(c/a)^4 \; \cot^2 \theta \; + \; 20}{[1+(c/a)^2]^4} \; \right]\;\;\frac{a^2 \; 
\Omega^3_0 \; \rho}{\mu \; Q} \;\;\;\;,
\label{rate}
\ee
where $\;\mu\;$ is the shear elastic modulus of the material, $\;Q\;$ is its 
quality factor, and 
\be
\Omega_0\;\equiv\;\frac{J}{I_3} \;\;\;\;
\label{kozel}
\ee
is a typical angular velocity. The above formula (\ref{rate}) shows that the 
major-inertia axis slows down its alignment for vanishing $\; \theta \;$, which
looks reasonable. Formula (\ref{rate}) differs by a factor of 2 from the 
appropriate formula in (Lazarian and Efroimsky 1998) because of an error in 
our preceding article (see Appendix for details).

It would be natural also to expect that the rate of alignment decays to zero 
for $\; c/a \;$ approaching unity. (For $\;c\,=\,a\;$, the body simply lacks 
a maximum-inertia axis.) One may as well expect to see not only a ``slow 
finish'' but also a ``slow start'' (so that $\;d\theta/dt\;$ vanish for 
$\;\theta\,\rightarrow\,\pi/2\;$): the maximum-inertia axis must be hesitant 
as to whether to start aligning along or opposite to the 
angular momentum. Still, if we look at (\ref{rate}), it will appear that 
$\; d \theta /dt \;$ remains nonvanishing for $\;c/a\;$ approaching unity, 
and that the major axis 
leaves the position $\; \theta = \pi/2 \;$ at a finite rate. Recall, 
however, that all the above machinery works only insofar as the 
rotation and precession are fast, while the alignment is slow: $\; 
{\dot{\theta}} \ll \omega \;$. 

The time needed for the maximal-inertia axis to shift toward 
alignment with the angular momentum is:
\begin{equation}
t \; \;  \; \; \equiv - \int_{\delta}^{\theta_0} \; 
\frac{d \theta}{d \theta/dt} \; \; \; .
\label{6.22}
\end{equation}
where 
$\theta_0$ is the initial angle ($\theta_0<\pi/2$), while a small but 
finite $\delta$ is introduced to avoid the ``slow-finish'' divergency. 
Each pair of values of $\theta_0$ and $\delta$ will give birth to one or 
another numerical factor of order unity, in the expression for the relaxation 
time. Calculations presented in Appendix B show that $t$ is not very 
sensitive to the choice of angle $\theta_0$ as long as this angle is not too 
small. (This weak dependence upon the initial angle is natural since the 
divergence emerges at small angles in the end of the motion.) {\it A 
particular 
choice of} $\delta$ {\it must be based on one's capability to recognise the 
precession by observational means.} In the case of ground-based photometric 
experiments, one studies the amplitude of 
the lightcurve variation (which in the first approximation is proportional to 
the variation in the cross-sectional area of the body). A typical observational
accuracy is {\it ~0.01 mag}; that is, deviations from one rotation to the next
less than 
{\it ~0.01 mag} cannot be confirmed to be real. From these assumptions, we 
arrive at some rough estimate of the minimal half-angle of the precession 
cone: $\;\delta\,\approx\,10^o\;.$ Radar can produce images with resolution 
as fine as a decameter, in an absolute reference frame, and can 
discern the dimensions and spin states of asteroids in detail.
In principle, with an optimum data set, an observation may reveal wobble 
with precession-cone half-angles of several degrees\footnote{Steven Ostro, 
private communication 1999}. For most radars ``several'' means: about five 
degrees\footnote{Scott Hudson, private communication 1999}. To be on the safe 
side, we shall take
\be
\delta\;=\;6^o\;\;\;\;,
\label{ugol}
\ee
though, with the NEAR spacecraft scheduled to rendezvous with (433) Eros in the
next year, nutations of only a degree or less may become detectable.

By plugging (\ref{rate}) into (\ref{6.22}) one can get the relaxation time $\;t
\;$ as a function of $\;\theta_0\;$ and $\;c/a\;$. A simple computation shows 
that $\;t\;$ is not particularly sensitive to the oblateness
\break
\begin{figure}
\centerline{\epsfxsize=3.5in\epsfbox{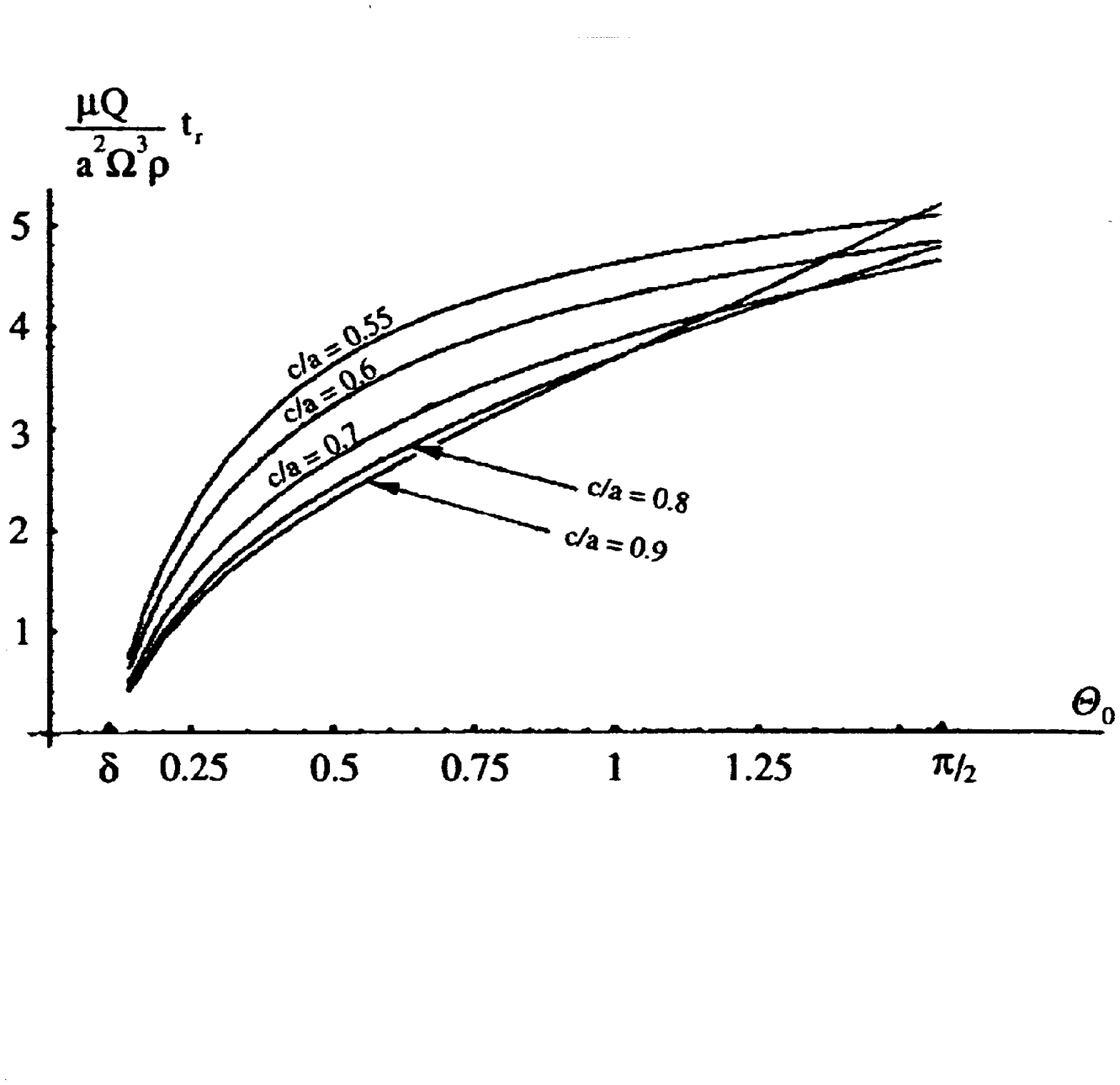}}
\bigskip
\caption{Comparison of the approximation (2.11) with the exact results for the 
relaxation time $\,t$. The exact calculation of $\,t\,$ (scaled by a factor of 
$\;\mu \, Q/a^2\,\Omega^3_0\,\rho\;$) was performed by combining formulae (2.9)
and (2.7) for various realistic values of the parameter $\;c/a\;$. The final 
value of the angle $\;\theta\;$ was assumed to be $\;\delta\,=\,6^0\;$, while 
its initial value was varying from $\;\delta\;$ through $\;\pi/2\;$.
}
\end{figure}
\noindent\\
parameter $\;c/a\;
$ when values of this parameter fall within the interval 0.5 - 0.9 (see Fig.1).
As we can see from the plot, the dependence $\;t(\theta_0)\;$ may be approximated, for $\;c/a\,=\,$0.5 - 0.9, by the following expression that does not bear 
any dependence upon $\;c/a\;$:
\begin{eqnarray}
t\;\;\simeq \;\;-\;2.80\;\theta^2_0\;+\;7.84\;\theta_0\;-\;0.79
\label{6.23}
\end{eqnarray}

This is a rough approximation. For the above value of $\;\delta\;$, it gives 
an error of about 100 \% for $\,\theta_0\,\stackrel{>}{\sim}\,\delta\,$ and an 
error of about 10 \% for $\,\theta_0\,>\,\pi/4\,$. Nevertheless it is more 
than acceptable for our purposes, for it enables one to get an idea as to what 
a typical value of the relaxation time could be.

\protect\section{Comparison with an earlier study}

We shall compare our result for the typical time of alignment 
with the estimate obtained in Burns \& Safronov (1973):
\begin{eqnarray}
t_{(B \; \& \;S)} \; \approx \; A \; \frac{\mu \; Q}{\rho \; a^2 \; 
\Omega^3_0} \; \; \; ,
\label{7.2}
\end{eqnarray}
the numerical factor A being about $100$ for 
bodies of small oblateness, and about $10$ for bodies of very irregular shapes.
In astrophysics we normally encounter the case of small oblateness 
(except perhaps some exotic species of cosmic-dust grains, called carbon 
flakes), so let us assume $\; 0.5\, \stackrel{<}{\sim}\,c/a \,\stackrel{<}{
\sim} \, 0.9 \,$.  Our formula will give:
\begin{eqnarray}
\nonumber
t_{(our \; result)}^{c \approx a} \; \approx \; (1\;-\;2) \; \frac{\mu \; Q}{
\rho \; 
a^2\; \Omega^3_0} \;\;\;\; for \;\;\;\; \theta_0\,\approx\,(2\;-\;3)\, \delta
\ea
\ba
t_{(our \; result)}^{c \approx a}\;\approx \;(3\;-\;4) \;\frac{\mu \; Q}{\rho 
\;a^2 \; \Omega^3_0} \; \; \; \; for \;\;\;\; \theta_0\,\approx\,\pi/4
\label{87.3}
\ea
\ba
\nonumber
t_{(our \; result)}^{c \approx a} \; \approx \; (4\;-\;5) \; \frac{\mu \; Q}{
\rho \; a^2 \; \Omega^3_0} \;\;\;\; for \;\;\;\; \theta_0\,\stackrel{<}{\sim}\,\pi/2
\end{eqnarray}
while the aforequoted formula of Burns and Safronov will read:
\begin{eqnarray}
t_{(B \; \& \; S)}^{c \approx a} \; \approx \; 100 \; 
\frac{\mu \; Q}{\rho \; a^2 \; \Omega^3_0} \; \; \; .
\label{77.4}
\end{eqnarray}
We see that in the preceding study the relaxation time was overestimated, for 
small initial angles, by two orders. In other words, the effectiveness of the 
inelastic-relaxation process was underestimated by two orders.  Even when the 
initial 
angle  is not too small ($\;\theta_0\;\sim\;\pi/4\;$), the underestimation will
be by a factor of 30 or so\footnote{In the case of body of a very irregular 
shape, the discrepancy between Burns and Safronov's result and our calculation
will be less, because in this case Burns and Safronov suggest a smaller 
value for the numerical factor A in (\ref{7.2}). (See formula (23) in Burns 
\& Safronov (1973), and a paragraph thereafter.)}. The three 
main reasons for this underestimate are the following. First, it is the 
contribution of the double-frequency 
mode missing in (Burns \& Safronov 1973). In the sum $\; (63 (c/a)^4\,\cot^2 \theta + 20) \;$ emerging in 
formula (\ref{rate}), the term $\; 63 (c/a)^4 \,\cot^2 \theta\;$ appears due 
to the principal mode, while the term $\; 20 \;$ appears due to the 
double-frequency mode. Thence, in the case of $\; c \;$ close to $\; a \;$ the 
double-frequency mode will give (after integration over $\theta$) a 
considerable input, while for $\; c \ll a\;$ (which is not a likely case for 
asteroids and comets, but may happen for cosmic dust) it provides an 
overwhelming contribution to the entire effect. The only case when the 
contribution of the second mode is irrelevant is the case of a small initial 
angle $\;\theta_0\;$. But in this case there exists another simple reason for 
 (\ref{7.2}) to give a large error compared to our analysis: (\ref{7.2}) simply
ignores any dependence upon the initial angle, and this is why it gives too 
long times for small angles. The third source of error in (\ref{7.2})  
is an assumption, accepted by Burns and Safronov, that the energy 
dissipation in the case of small oblateness is predominantly due to
the bulge flexing. It seems that this simplifying assumption does not
work well. 

\protect\section{Dynamics of Prolate Bodies}

At first glance, the dynamics of a freely-spinning prolate body obeys
the same principles as the dynamics of an oblate one: the axis of
maximum inertia will tend to align itself parallel to the angular momentum. 
If we model a prolate body with a symmetric top, it will be once again 
convenient to choose it be a prism of dimensions $\; 2a \times 2a \times 2c 
\;$, though this time half-size $\, c \,$ is larger than $\, a \,$, and 
thereby $\; I_3 \; = \; I_2 \; > \; I_1 \;$. Then all our calculations 
{\textit{formally}} remain in force, except that the right-hand side in 
(\ref{6.21}) changes its 
sign\footnote{For details see (Lazarian and Efroimsky 1999)}, provided we keep
the notation $\theta$ for the angle between $\; \mathbf{J} \;$ and the 
body-frame axis 3 (parallel to dimension $\; 2c \;$).
This brings an illusion that axis 3 tends to stand orthogonal to $\mathbf{J}$, 
which at first glance appears natural since axis 3 is now not the 
maximum-inertia but the minimum-inertia axis.

Unfortunately, this extrapolation of the oblate-body description to a 
prolate-body case is of absolutely no practical relevance. The problem
is that an arbitrarily small difference between $\; I_3 \;$ and $\; I_2 \;$ 
will entail an entirely new character of precession (Synge \& Griffith
1959). For the first time this topic, in the context of asteroid and
comet precession, was addressed by Black et al (1999). We are planning to 
dwell on the subject comprehensively in our next article. 

\protect\section{Numbers}

Values of the involved parameters may depend both upon the temperature of the 
body and the wobble frequency. Evidently, the temperature-, pressure- and 
frequency-caused variations of the density $\; \rho \;$ are tiny and may be
neglected. This way, we can use the (static) densities appropriate to the 
room temperature and pressure: $\rho^{(silicate)} \; \approx \; 2500 \; kg \; 
m^{-3} \,$ and $\rho^{(carbon)} \; \approx \; 2000 \; kg \; m^{-3}$. 

As for the adiabatic shear modulus $\; \mu \;$, tables of 
physical quantities would provide its values at room temperature and 
atmospheric pressure, and for quasistatic regimes solely. As for the possible 
frequency-related effects in materials (the so-called ultrasonic attenuation),
these become noticable only at frequencies higher than $\; 10^8 \; Hz \;$ 
(see section 17.7 in Nowick and Berry 1972). Another fortunate circumstance is
that the pressure-dependence of the elastic moduli is known to be weak 
(Ahrens 1995). Besides, the elastic moduli of solids are known to be 
insensitive to temperature variations, as long as these variations are far 
enough from the melting point. The value of $\; \mu \;$ may increase by 
several percent when the temperature drops from room temperature to $\;
10 \, K \;$. Dislocations don't affect the elastic moduli either. Solute 
elements have very little effect on moduli in quantities up to a few 
percent. Besides, the moduli vary linearly 
with substitutional impurities (in which the atoms of the impurity replace 
those of the hosts). However hydrogen is not like that: it enters the 
interstices between the atoms of the host, and has marginal effect on the 
modulus. As for the role of the possible porosity, the elastic moduli scale 
as the square of the relative density. For porosities up to about $20\;\%~\,$,
 this is not of much relevance for our estimates\footnote{We express deep 
thanks to Michael Aziz and Michael Ashby for their consultations on these 
topics.}.

According to Ryan and Blevins (1987), for both carbonaceous and silicate rocks
one may take the shear-modulus value  $\mu \; \approx \; 10^{10} \; Pa \; \;$.
For the dirty ice Peale and Lissauer (1989) suggest,  for Halley's 
comet, $\;\mu \; \approx \; 10^{10} \; dyne/cm^2 \; = \; 10^{9}\; Pa \;$ while 
$\; Q \; < \; 100 \;$.

A proper choice of the values of the $\;Q-$factor for asteroids is to be the 
most problematic subject. As well known from seismology, the $\; Q-$factor bears 
a pronounced dependence upon: the chemical composition, graining, frequency, 
temperature, and confining pressure. It 
is, above all, a steep function of the humidity which presumably affects the 
interaction between grains. The  $\; Q-$factor is less sensitive to the
porosity (unless the latter is very high); but it greatly depends upon the
amount and structure of cracks, and generally upon the mechanical nature of 
the aggregate. Whether comets and asteroids are loose aggregates or solid 
chunks remains unknown. The currently available experimental data are 
controversial. On the one hand, it is a well-established fact that comets 
sometimes get shattered by tidal forces. Namely, Shoemaker-Levy 9
broke into 21 pieces on the perijove preceding the impact 
(Marsden 1993)\footnote{There is also a strong evidence in favour of Comet 
Brooks 2 having been rent into pieces by the jovian gravity in 1886 
(Sekanin 1982). One may as well mention the disintegration of Comet West in 1976 
(Melosh and Schenk 1993), though we would rather decline this argument because
the comet was too close to the Sun, and was therefore warmed up. Catenae of
aligned craters found on the lunar surface may become another possible evidence
of comets being prone to shattering. The strongest argument in favour of the 
rubble-pile hypothesis comes from the recent discovery of such catenae on 
Callisto and Ganymede (Melosh and Schenk 1993).}. It seems that {\it {at least some 
comets are loosely connected aggregates, though we are unsure if all  
comets are like this}}.  

As for the asteroids, the question is yet open, and 
the low density of asteroid 253 Mathilde (about 1.2 $\; g/cm^3$) may be 
either interpreted in terms of the rubble-pile hypothesis (Harris 1998), or 
put down to Mathilde being perhaps mineralogically akin to low-density 
carbonaceous chondrites, or be explained by a very 
high porosity\footnote{It should be emphasised at this point that in our 
opinion high porosity of a material does not necessarily imply this material 
being a rubble pile. There exist also an opposite viewpoint: according to 
Steven Ostro and Alan Harris (private communication 1999), high porosities 
(tens of percent) indicate that much of the body is unconsolidated.}. In our 
opinion, the sharply-defined craters on the surfaces of some asteroids 
witness {\bf{against}} the application of the rubble-pile hypothesis to 
asteroids
\footnote{This is just our opinion. According to Steven Ostro's opinion  
(private communication 1999), the craters on Mathilde comparable to the object's 
radius can only be made in a rubble-pile asteroid.}. And of course, we 
should mention here Vesta as a reliable example of an asteroid being a 
solid body of a structure common for terrestial planets: Hubble images of 
Vesta have revealed basaltic regions of solidified lava flows, as well as a 
deep impact basin exposing solidified mantle. Thus we would say 
that {\it{at least some asteroids are well connected solid chunks, though we
are uncertain whether this is true for all asteroids}}\footnote{Currently 
available radar and optical data establish that the ~ 30-meter object 
$\;1998\;KY26\;$ is monolithic, and suggest that at least a few larger objects 
are also monolithic (Ostro 1999).}.

In the sequel, hence, we shall assume that the body is not a loosely 
connected aggregare but a solid rock (possibly porous but nevertheless solid 
and well-connected), and shall try to employ some knowledge available on  
attenuation in the terrestial and lunar crusts. 

We are in need of the values of the quality factors for silicate and 
carbonateous rocks. We need these at the temperatures from several K to 
dozens of K,  , 
zero confining pressure, frequencies appropriate to asteroid precession 
($10^{-6} \; - \; 10^{-4} \; Hz$), and (presumably!) complete lack of moisture.

Much data on the behaviour of 
$\;Q-$factors is presented in the seismological literature. Almost all
of these measurements have been made under high temperatures (from
several hundred up to 1500 Celsius), high confining 
pressures (up to dozens of MPa), and unavoidably in the presence of 
humidity. Moreover, the frequencies were typically within the range from  
dozens of $kHz$ up to several $MHz$. Only a very limited number of measurements
have been performed at room pressure and temperature, while 
no experiments at all have been made thus far with rocks at low temperatures
(dozens of $\;K\;$). The information about the role of humidity is 
extremely limited. Worst of all, only a few experiments were made with rocks at
the lowest seismological frequences ($10^{-3} \; - \; 10^{-1} \; Hz$), and 
none at frequencies between $\; 10^{-6} \;$ and $\; 10^{-2} \; Hz$,  
though some indirectly achieved data are available (see Burns 1977, Burns 
1986, Lambeck 1980, and references therein).  

It was shown by Tittman et al. (1976) that the $\; Q-$factor of about $\; 60 
\;$ measured under ambient conditions on an as-received lunar basalt was 
progressively increased ultimately to about $\; 3300 \;$ as a result of 
outgassing under hard vacuum. The latter number will be our starting point. 
The measurements were performed by Tittman et al. at $\; 20 \; kHz \;$ 
frequency, room temperature and no confining pressure. How might we estimate
the values of $\; Q \;$ for the lunar basalt, appropriate to the
lowest frequencies and temperatures? 

As for the frequency-dependence, it is a long-established fact (e.g., 
Jackson 1986, Karato 1998) that $\; \; Q \; \sim \omega^{\alpha} \;$ with $\;
 \alpha \;$ around 0.25. This dependence reliably holds for all rocks 
within a remarkably broad band of frequencies: from hundreds of $\; kHz \;$ 
down to $\;10^{-1} Hz$. Very limited experimental data are avaliable for 
frequencies down to $\;10^{-3} Hz$, and none below this
threshold. Keep in mind that $\; \alpha \;$ being close to 0.25
holds well only at temperatures of several hundred Celsius and higher, 
while at lower temperatures $\; \alpha \;$ typically decreases to 0.1
and less.

As regards the temperature-dependence, there is no consensus on this point in
the geological literature. Some authors (Jackson 1986) use a simple rule:
\be
\; Q \; \sim \; \omega^{\alpha} \; \exp(A^*/RT) \; \;\;,     
\label{wrong}
\ee
$\;A^*\;$ being the apparent activation energy. A more refined 
treatment takes into account the interconnection between the 
frequency- and temperature-dependences\footnote{The authors are grateful to 
Shun-ichiro Karato who drew our attention to this connection.}. Briefly 
speaking, since the quality factor is dimensionless, it must retain this 
property despite the exponential frequency-dependence. This may be achieved 
only in the 
case that $\; Q \;$ is a function not of the frequency {\it{per se}} but of 
a dimensionless product of the frequency by the typical time of 
defect displacement. The latter exponentially depends upon the activation 
energy, so that the resulting dependence will read: 
\be
\; Q \; \sim \; \left[ \omega \; \exp(A^*/RT) \protect\right]^{\alpha} \;\;\;, 
\label{right}
\ee
where $\; A^* \; $ may vary from 150 - 200 $\; kJ/mol \;$ (for dunite and 
polycristalline forsterite) up to 450 $\; kJ/mol \;$ (for olivine). 
This interconnection between the frequency- and temperature-dependences 
tells us that whenever we lack a pronounced frequency-dependence, the 
temperature-dependence is absent too. It is known, for example 
(Brennan 1981) that at room temperature and pressure, at low 
frequencies ($10^{-3} \; - \; 1 \; Hz$) the shear $\; Q-$factor is 
almost frequency-independent for granites and (except 
some specific peak of attenuation, that makes $Q$ increase twice) for basalts.
It means that within this range of frequencies $\; \alpha \;$ is 
small (like 0.1, or so), and $\;Q\;$ may be assumed almost temperature-independent too. 

Presumably, the shear Q-factor, reaching several thousand at $\; 20 \;kHz 
\;$, descends, in accordance with (\ref{right}), to several hundred when  
the frequency decreases to several $\; Hz$. Within this band of frequencies, 
we should use the power $\;\alpha\,\approx\,0.25\;$, as well known from 
seismology. When we go to lower 
frequencies (from several $\; Hz$ to the desirable $10^{-6} \; - \; 10^{-4} \;
Hz$), the Q-factor will descend at a slower pace: it will obey (\ref{right}) 
with $\;\alpha\,<\,0.1\;$. Low values of $\; \alpha \;$ at low  
frequencies are mentioned in Lambeck (1980) and in Lambeck 
(1988)\footnote{See also Knopoff (1963) where a very slow and smooth frequency-dependence of $Q$ at 
low frequencies is pointed out.}. The book by Lambeck containes much material 
on the $Q$-factor of the Earth. Unfortunately, we cannot employ the numbers 
that he suggests, because in his book the quality factor is defined for the 
Earth as a whole. Physically, there is a considerable difference between the 
$Q$-factors emerging in different circumstances, like for example, between 
the effective tidal 
$Q$-factor\footnote{A comprehensive study of the effective tidal $Q$-factors 
of the planets was performed by Goldreich and Soter (1966)}
and the $Q$-factor of the Chandler wobble). In regard to the latter, Lambeck 
(1988) refers, on page 552, to Okubo (1982) who suggested that for the 
Chandler 
wobble $\;50\,<\,Q\,<\,100\;$. Once again, this is a value for the Earth as a 
whole, with its viscous layers, etc. We cannot afford using these numbers 
for a fully solid asteroid.
               
Brennan (1981) suggests for the shear $\; Q-$factor the following 
values\footnote{Brennan mentions the decrease of $Q$ with humidity, but 
unfortunately does not explain how his specimens were dried.}: 
$Q^{(granite)}_{(shear)} \; \approx \; 250  \; $, $\; Q^{(basalt)}_{(shear)} 
\; \approx \; 500 \;$.
It would be tempting to borrow these values\footnote{Recall that these
data were obtained by Brennan for strain amplitudes within the linearity 
range. Our case is exactly of this sort since the typical strain in a
tumbling body will be about $\; \sigma/ \mu \; \approx \; \rho \Omega^2
\; a^2/ \mu \,.$ For the size $\; a \; \approx \; 1.5 \, \times \,
10^4 \, m$ and frequency not exceeding $\; 10^{-4} Hz$, this strain is
less than $\; 10^{-6} \,$ which is a critical threshold for linearity.}, if 
not for one circumstance: 
as is well known, absorption of only several monolayers of a saturant may 
dramatically decrease the quality factor. We have already mentioned
this in respect to moisture, but the fact is that this holds also for 
some other saturants\footnote{like, for example, ethanol (Clark et al. 1980)}. 
Since the asteroid material may be well saturated with hydrogen (and possibly 
with some other gases), its $\,Q$-factor may be much affected. 

It may be good to perform experiments, both on carbonaceous and
silicaceous rocks, at low frequencies and temperatures, 
and with a variety of combinations of the possible saturants. These 
experiments should give us the values for
both shear and bulk quality factors. The current lack of experimental data 
gives us no choice but to start with the value 3300 obtained by Tittman for 
thoroughly degassed basalts, and then to use formula (\ref{right}). This will 
give us, at $\;T\,=\,20\;K\;$ and $\;\omega\,=\,10^{-5}\;Hz\;$:
\be
Q^{(basalt)}_{(shear)} \; \approx \; 150 
\label{formula}
\ee 
This value of the shear $Q$-factor\footnote{The bulk $Q$-factor may differ 
from the shear one. In our case we have a sophisticated deformation picture 
that includes both torsional and longitudinal displacements. Simply from 
looking at the expressions for stresses we can see that torsion will 
dominate. Hence the effective $Q$-factor must be close to the shear 
$Q$-factor  value.} for granites and basalts differs from the one chosen in 
Burns and Safronov (1979) only by a factor of 2. For carbonaceous materials 
$\; Q \;$ must be surely much less than that of silicates, due to weaker 
chemical bonds. So for carbonaceous rocks we shall choose the following upper 
boundary:
\be
Q^{(carb)} \; < \; 100 \; \; \; ,
\label{9.2}
\ee
though this boundary is somewhat arbitrary, and is probably still too high.

Terrestial geophysics has not yet given us a reliable handle on the acoustic 
properties of asteroids and comets, though the necessary quality
factors can be obtained by a modification of the existing testing technique. 
Namely, the frequencies must be within the interval $\; 10^{-6} \; - \;
10^{-4} \; Hz$, the temperature must be from about $\;K \;$ to dozens of 
$\;K\;$. The specimens 
must be well outgassed and the measurements must be performed under a 
high vacuum that would mimic the real interplanetary environment.
It would be most important to study specimens exposed to a variety of 
saturants (hydrogen, first of all). The specimens must include both  
silicate and carbonaceous rocks, and artificially or naturally 
prepared samples of dirty ice.

Twenty two years ago J.A. Burns (1977) wrote: ``Further experimental and 
theoretical work for real materials is sorely needed if we are to ever trace 
the history of the natural satellites.'' This plea is even more relevant 
today, and the ramifications of such work will nowadays be even broader: 
these studies are to become our key not only to satellite but also to 
asteroid studies. 
 
\section{Particular example: Asteroid 4179 Toutatis}

The asteroid 4179 Toutatis is a slowly rotating body of size about 2
 $km$, that was ``imaged'' in radar   
by Ostro et al. (1993), (1995), (1999), Hudson and Ostro 1995, 
Scheeres et al. 1998. It is of S-type, analogous to 
stony irons or ordinary chondrites but certainly not to carbonaceous 
chondrites (Ostro et al. 1999). It has a rotation period 
$ \; \tau \; = \; 7.5 \; days 
\;$ (so that $\Omega_0 \; = \; 2 \pi/\tau \; \approx \; 9.7 \times 10^{-6} \; 
Hz$).

Our analysis was aimed at oblate bodies and, rigorously speaking, it is 
unapplicable to prolate or triaxial rotators. Still, while a similar study 
for triaxial rotators in on the way, let us try to apply our formulae, as 
a zeroth approximation, to Toutatis. Suppose that some tidal interaction has 
forced this asteroid to 
precess with a small precession-cone half angle $\;\theta_0\;$. Let it be, for 
example, twice or thrice the minimally recognisable half-angle: $\;\theta_0\;=
\;(2\,-\,3)\,\delta\;\sim\; 12^o\;$. Then Fig.1 (or our formula (\ref{87.3}) 
based on it) will give, for $\; Q \, = \, 150 \;$, $\; \mu \, = \, 10^{11} \, 
dyne/cm^2 \, = \, 10^{10} \; Pa \;$, and $\; \rho \,= \, 2000 \; kg/m^3$, $\;
c/a\,\stackrel{>}{\sim}\,1/2\;$:
\be
t \; \stackrel{<}{\sim} \; 2.5 \; \frac{\mu\;Q}{\rho\,a^2\,\Omega^3_0}\;=\;0.5
\;\times 10^{18} \; s \; \approx \; 1.6 \; \times
\; 10^{10} \; years 
\label{10.1}
\ee
which is 800 times less than the estimate that would come 
from Burns \& Safronov's (1973) treatment. Factor 800 emerges, first, due 
to the difference by a factor of 40 between our formula (\ref{10.1}) and 
Burns \& Safronov's formula (\ref{77.4}); and second, due to the difference in 
choice of the quality-factor and shear-modulus values. Our choice is: 
$\; \mu Q \; = \; 1.5 \; \times \; 10^{13} \; dyne/cm^2 \; = 
\; 1.5 \; \times \; 10^{12} \; Pa \;$,
while Burns \& Safronov suggested  $\; \mu Q \; = \; 3 \; \times
\; 10^{14} \; dyne/cm^2 \; = \; 3 \; \times \; 10^{13} \; Pa \;$. 
We suggest values lower than in Burns \& Safronov (1973) for the
reasons explained above.

Harris (1994) goes even further: he suggests to use data 
measured for Phobos:  $\; \mu Q \; = \; 5 \; \times \; 10^{12} \; dyne/cm^2 \;
 = \; 5 \; \times \; 10^{11} \; Pa \;$ which is almost 
two orders of magnitude less than that in Burns \& Safronov (1979).
Therefore if we adopt the values proposed by Harris (1994), our
estimate (\ref{10.1}) will be further decreased by 3 times, and will equal to 
$\; 5 \; \times \; 10^{9} \; years \;$, which is $\;10^3\;$  
times less than the result that would follow from Burns \& Safronov's 
treatment. 

Even though Harris chose a lower value for $\mu Q$ than we did, our
calculation will give a lower value for the relaxation time, due to
the difference in numerical factors in our formula (\ref{87.3}) and
Burns \& Safronov's formula (\ref{77.4}). (Actually, this difference 
is even larger than it should be, because Harris, applying Burns \& Safronov's 
formula, took the overall numerical factor to be not 10 but 30.) This would 
mean that
the number of asteroids that are suspected of tumbling (see Fig.~1 in 
Harris (1994)) should be decreased.

How certain is this conclusion? It follows from our discussion in the
previous section that at least for solid asteroids the $\mu Q$ factor
obtained for Phobos in Yoder (1982) may be an
underestimate. Observations of asteroid precession 
should provide us with insight into their mechanical properties.
The comparison of these with the properties of laboratory-studied
materials should provide us with more understanding of the composition
and inner structure of asteroids.

While studying wobbling, it is important to bear in mind two points.
First of all, nutations at small angle can proceed for much longer
times than we estimated above. This is the consequence of
the ``slow finish'' condition that we discussed in section~III.
With the improvement of the radar measurements, we expect to detect
more asteroids wobbling at small angles. Therefore high-accuracy imaging of
asteroids may reveal nutations above the upper curve in Fig.~1 in Harris 
(1994).
Second, large-amplitude or/and rapid tumbling may be
suppressed quicker than we estimated  
as the strains become nonlinear and the dissipation 
increases\footnote{As  already mentioned
above, the linearity threshold is, for different materials, typically between 
$\; 10^{-6} \;$ and $\; 3 \, \times \, 10^{-6}$. A body of density 
$\; \rho \, = \, 2000 \; kg/m^3 \;$, size     
$\; a \, = \, 20 \; km \;$, rotating with a 
period of $\, \tau \, = \, 6 \,$ hours (i.e., with $\, \Omega_0 \, =
\, 2 \pi/\tau \, = \, 3 \, \times \, 10^{-4} \, Hz$) will
experience in the cause of precession, in the regions farthest from its 
centre of rotation, strain of amplitude exceeding the linearity threshold.}.

\section{Conclusions}

1. The dissipation at the double frequency of precession makes an importantm, 
sometimes decisive, contribution into the process of inelastic relaxation.

2. Small-amplitude wobbling can proceed for long periods of time, while
the large-amplitude precession gets damped to smaller amplitudes much faster. 
This is a consequence of the dependence of the relaxation rate upon
the angle of precession. In reality, the finite resolution of radar-generated 
images makes us take into consideration only wobble with precession-cone 
half angles no less than about $\;5\,-\,6^o\;$. (Though observations made by 
spacecrafts will, hopefully, improve the resolution up to one degree or so.)

3. Neglection of the above two circumstances, along with acceptance of an 
 assumption (unjustified, in our opinion) that the tidal-bulge flexing plays a 
decisive role in the inelastic relaxation, lead our predecessors to a great 
underestimate of the effectiveness of the inelastic dissipation mechanism. 
(By two orders, for small initial angles, and by a factor of 20 - 30 for large 
angles.)

4. Solid asteroids are likely to have $Q-$factors about one hundred, or 
possibly even less, but further laboratory research is required. 
Earth seismological data may poorly represent acoustic properties of 
asteroids and comets 
as even a tiny trace of moisture (several monolayers) substantially 
alters the quality factor. Laboratory data on  $\; 10^{-6} \; 
- \; 10^{-4} \; Hz \;$ oscillations of de-moisturised silicate and
carbonaceous rocks at temperatures within the range from several $\;K\;$ 
to several dozens $\;K\;$ are badly needed. 

5. Potentially, wobbling may provide valuable information on the
composition and structure of asteroids, and on their recent history of 
external impacts.

6. Since the inelastic dissipation turns out to be a far more effective process
than believed previously, the number of asteroids expected to wobble with a 
large angle of the precession cone is expected to be lower (than  
the predictions made in (Harris 1994).)

{\bf Acknowledgements}  

The authors are grateful to Eric Heller for encouragement, and to Alan Harris 
and Steven Ostro for stimulating discussions  
that helped us to considerably improve the article.  
M.E. would like to acknowledge very useful information on the 
properties of materials, provided by Michael Aziz and Michael Ashby, and 
a consultation on radar accuracy, by Scott Hudson.  
A.L. acknowledges discussions with Bruce Draine, and a couple of suggestions 
by Scott Tremaine. 
The work of A.L. was supported by NASA grant NAG5-2858. 
This paper would never have been completed without a comprehensive tutorial 
on the Q-factor, so kindly offered to us by Shun-ichiro Karato: his 
consultation gave a second breath to our project.  

\pagebreak

\appendix

\section{Alignment of the maximal-inertia axis  
towards the angular-momentum vector \label{A}}

If we model the freely rotating oblate body by a prism of sizes
$2a \times 2a \times 2c$, ($c < a$) and density $\;\rho\;$, the time-dependent
 stresses will read (Lazarian and 
Efroimsky 1998):
\be
\sigma_{xx} = \frac{\rho \Omega_\perp^2}{4}  
(x^2 - a^2) \; \cos 2 {\omega} t  \; \; , \; \; \;  
\sigma_{yy} = - \frac{\rho \Omega_\perp^2}{4} (y^2 - a^2) \;  
\cos 2 {\omega} t \; \; , \; \; \;  
\sigma_{zz} = 0 
\label{4.5}
\ee
\be
\sigma_{xy} \; \; = \; \; \frac{\rho}{4} \; \Omega_\perp^2 \; 
(x^2 \; \; + \; \; y^2 \; \; - \; \; 2a^2)
\; \; \sin \; 2 {\omega} t \; \; \; ,
\label{4.6}
\ee
\be
\sigma_{xz} \; \; = \; \; \frac{\rho}{2} \; {\Omega_\perp} \; {\Omega_3} \; 
\left[ \; h \; (z^2 \; - \; c^2)
\; \; + \; \; (2 \; - \; h) \; (x^2 \; - \; a^2) \; \right] \; \; 
\cos \; {\omega} t \; \; \; ,
\label{4.7}
\ee
\be
\sigma_{yz} \; \; = \; \; \frac{\rho}{2} \; {\Omega_\perp} \; {\Omega_3} \; 
\left[ \; h \; (z^2 \; - \; c^2)
 \; \; + \; \; (2 \; - \; h) \; (y^2 \; - \; a^2) \; \right] \; \; 
\sin \; {\omega} t \; \; \; .
\label{4.8}
\ee
As expected, not only the principal frequency but also the second mode shows up
in $\sigma_{ij}$.

The moment of inertia $\; I_3 \;$ and the parameter $\; h \;$ read:
\be
I_3 \;  = \; \frac{16}{3} \; \rho \; a^4 \; c\;\;\;\;\;,\;\;\;\;\;\;\;\;
h \; \; \equiv \; \; \frac{I_3}{I} \; \; = \; \; \frac{2}{1 \; + \; (c/a)^2}  
\; \; \; \; \; . \; \; \; 
\label{4.4}
\ee
According to (\ref{om4}),
\be
\Omega_3 \; \; = \; \; \Omega_0 \; \cos \theta 
\; \; \; \; \; , \; \; \; \; \; \; \; \; \; \; \; \; 
\Omega_{\perp} \; \; = \; \; \Omega_0 \; h \; \sin \theta \; \; \; ,
\label{4.13}
\ee
$\Omega_0 \equiv  J/I_3$ being the typical angular velocity 
of a body. We shall compute the 
strain tensor in order to estimate the maximal elastic energy 
$\, W \,$ stored in the body. Then, we shall estimate the dissipation
rate as $\, -\omega W/Q \,$,  $\; Q \;$ being the quality factor
of the material.

The kinetic energy of an oblate spinning body reads, according to
(\ref{2.6}), (\ref{rotation}), and (\ref{om4}):
\begin{eqnarray}
E_{rot} \; = \; \frac{1}{2} \;
[I \; \Omega_{\perp}^2 \; + \; I_3 \; {\Omega_3}^2] \;
= \; \frac{1}{2} \; \; \left[ \; \frac{1}{I} \; \; \sin^2 \; \theta \; \; + \; \; 
\frac{1}{I_3} \; \; \cos^2 \; \theta \; \right] \; J^2
\label{5.1}
\end{eqnarray}
so that 
\be
\frac{dE_{rot}}{d\theta} \; \; = \; \; 
\frac{J^2}{I_3} \; (h \; - \; 1) \; \; \sin \; \theta \; \; \cos \; \theta 
\; \; =  \; \; \omega \; J \; \; \sin \; \theta \; \; \; .
\label{5.2}
\ee
The rotational energy changes via the inelastic dissipation:
\be
\dot{E}_{rot} \; = \; \dot{W} \; \; \; ,
\label{5.3}
\ee
$\; W \;$ standing for the elastic energy of the body. Then the rate of 
alignment is:
\be
\frac{d\theta}{dt} \; = \; \left(\frac{dE_{rot}}{d\theta}\right)^{-1} 
\frac{dE_{rot}}{dt} \; = \; \left( \omega \; J \; \; \sin \; \theta 
\right)^{-1} \; \dot{W}
\label{5.4}
\ee
where
\be
\dot{W} \; = \; \dot{W^{({\omega})}} \; + \; \dot{W^{(2{\omega})}} \; = \;-\;
\omega \; \frac{W_0^{({\omega})}}{Q^{({\omega})}} \; - \; 2 \; \omega \; 
\frac{W_0^{({2\omega})}}{Q^{({2\omega})}} \; \approx \; \frac{2\,\omega}{Q} \; 
\left\{ W^{({\omega})} \; + \; 2 W^{({2\omega})} \protect\right\} \; \; \; , 
\label{5.5}
\ee
the quality factor being almost frequency-independent. In reality, it 
certainly does depend upon $\omega$, but the dependence is
very smooth and may be neglected for frequencies differing by a factor of two.
(It should be taken into account within frequency spans of several orders.)
In (\ref{5.5}) $W_0^{\omega}\,,\;W_0^{2\omega}$ denote the amplitudes of the 
elastic energies associated with vibrations at the modes $\omega $ and $2
\omega $, while $W^{\omega}\,=\,2\,W_0^{\omega}\,$ and $\,W^{2\omega}\,=\,2\,
W_0^{2\omega}$ stand for the averages. In our preceding publication (Lazarian 
and Efroimsky 1998) we missed the coefficient 2 between  $W_0^{(...)}$ and 
$W^{(...)}$. 

Now the question is: how to calculate $\; W^{({\omega})} \; + \; 2 
W^{({2\omega})} \;$? At low temperatures the bodies manifest,
for small\footnote{The term ``small deformation'' means not exceeding the 
elastic limit. According to Brennan (1981), such deformations  
cause strain amplitudes not exceeding (for most materials) $\; (1 \, - \, 3) \times 10^{-6}$} 
displacements (like, say, sound) no viscosity: $\; \; \omega \eta \, \sim \, 
\omega \zeta \; \ll \;  \mu \, \sim \, K \;$. Hence the stress tensor will be 
approximated, to a high accuracy, by its elastic part:
\be
\epsilon_{ij} \; \; = \; \; \delta_{ij} \; \; \frac{Tr \; 
\sigma}{9 \; K} \; \; + \; \; 
\left( \; \sigma_{\it{ij}} \; \; - \; \; \frac{1}{3} \; \; \delta_{ij} 
\; \; Tr \; \sigma \right) \; \frac{1}{2 \; \mu} \; \; \; ~~~,
\label{6.1}
\ee
$ K \,$ and $\, \mu \,$ being the bulk and shear moduli. The elastic 
energy stored in a unit volume is
\ba
dW/dV \; = \; \frac{1}{2} \; \epsilon_{ij} \; \sigma_{ij} \; =\;\frac{1}{4} \; 
\left\{ \left(\frac{2 \; \mu}{9\; K} \; - \; \frac{1}{3} \protect\right) \, 
\left(Tr \; \sigma \protect\right)^2 \; + \; \sigma_{ij} \,  \sigma_{ij} 
\protect\right\} \;\;\;\;.
\label{6.2}
\ea 
Anticipating different rates of dissipation in the two modes, we  
split $dW/dV$ into parts: 
\be
dW/dV \; = \; dW^{(\omega)}/dV \; + \; dW^{(2\omega)}/dV
\label{6.3}
\ee
According to (\ref{5.5}), what we need is the sum
\ba
W^{({\omega})} \; + \; 2 W^{({2\omega})} \; = \; \int^{a}_{-a} \, dx \; 
\int^{a}_{-a} \, dy \; \int^{c}_{-c} \, dz \; \left\{ dW^{({\omega})}/dV \; + 
\; 2 \; dW^{({2\omega})}/dV  \protect\right\} 
\label{6.14}
\ea
To compute this sum, one has to plug the expressions for stresses (\ref{4.5}) -
(\ref{4.8}) into (\ref{6.2}), then to split the result in accordance with 
(\ref{6.3}), and to build up the sum under the integral in
(\ref{6.14}). These calculations (presented in (Lazarian and Efroimsky 
1999)) yield: 
\be
W^{({\omega})} \, + \, 2 \, W^{({2\omega})} \, = \, \frac{3^2 \, 2^{-6} \, 
(63 \, (c/a)^4 \, \tan^{-2} \theta \,+ \, 20)}{\mu} \;
\frac{a^{-2} \; c^{-1}}{ \left[1+(c/a)^2  \protect\right]^4} \; 
\left( \frac{I_3 \, \Omega_0^2}{2} \protect\right)^2 \; \sin^4 \theta \;\;\;
\label{6.18}
\ee
where $\; \Omega_0 \, \equiv \, J/I_3 \;$ is a typical angular velocity. 
Eventually, (\ref{5.4}), (\ref{5.5}) and (\ref{6.18}) will yield:
\be
d \theta/dt \; = \; - \; \frac{3^2 \; 2^{-6} \; 
(63 \; (c/a)^4 \; \tan^{-2} \theta \; + \; 20)}{\mu \; Q \; J} \;
\frac{a^{-2} \; c^{-1}}{ \left[1+(c/a)^2  \protect\right]^4} \; 
\left(  \frac{I_3 \; \Omega_0^2}{2}     \protect\right)^2 \; \sin^3 \theta
\label{6.19}
\ee
According to (\ref{4.4}) and (\ref{kozel}),
\be
\frac{1}{J} \; \left( \frac{I_3 \, \Omega^2_0}{2} \protect\right)^2 \; = \; 
\frac{4}{3} \; \Omega_0^3 \; \rho \;a^4 \; c
\label{6.20}
\ee
Substitution of the latter in the former gives the final expression for 
the alignment rate:
\be
d \theta/dt \; = \; - \; \frac{3}{2^4} \; \sin^3 \theta \; \; \frac{63 \; 
(c/a)^4 \; \cot^2 \theta \; + \; 20}{[1+(c/a)^2]^4} \; \; \frac{a^2 \; 
\Omega^3_0 \; \rho}{\mu \; Q}
\label{6.21}
\ee

\pagebreak

\pagebreak

\begin{thebibliography}{99}

\bibitem{} Ahrens, T.J. 1995 Ed., {\emph{Mineral Physics \& Crystallography. 
           A Handbook of Physical Constants.}} American Geophysical Union, 
           Washington DC 
\bibitem{} Black, G.J, P.D.Nicholson, W.Bottke, Joseph A.Burns, \& Allan W.  
           Harris 1999, "On a Possible Rotation State of (433) Eros''
           - Icarus, Vol. 140, p. 239
\bibitem{} Brennan, B.J. 1981, in: Anelasticity in the Earth (F.D.Stacey, 
           M.S.Paterson and A.Nicolas, Editors), Geodynamics Series
           4, AGU, Washington.
\bibitem{} Burns, J. A. and Safronov, V.S. 1973, MNRAS, 165, p. 403
\bibitem{} Burns, J. A. 1977, in: {\emph{Planetary Satellites}}, 
           J.A.Burns, Ed., University of Arizona Press, Tucson
\bibitem{} Burns, Joseph A. 1986, in: {\emph{Satellites}}, 
           J.A.Burns, Ed., University of Arizona Press, Tucson  
\bibitem{} Clark V.A., B.R.Tittman and T.W. Spencer 1980, J. Geophys. Res.,
           85, p. 5190
\bibitem{} Fowles, Grant R., and George L. Cassiday 1986 {\emph{Analytical 
           Mechanics.}} Harcourt Brace \& Co, Orlando, FL
\bibitem{} Goldreich, Peter, \& Steven Soter 1965, Icarus, Vol. 5, p. 375
\bibitem{} Harris, Alan W. 1994, Icarus, 107, p. 209
\bibitem{} Harris, Alan W. 1998, Nature, Vol. 393, p. 418
\bibitem{} Hudson, R.S., and S. J. Ostro 1995,  Science 270, 84-86 
\bibitem{} Jackson, Ian 1986, in: {\emph{Mineral and Rock Deformation. 
           Laboratory Studies}}, American Geophysical Union, Washington DC
\bibitem{} Karato, Shun-ichiro 1998, Pure and Applied Geophysics, in press
\bibitem{} Knopoff, L. 1963, Reviews of Geophysics, Vol.2, p. 625
\bibitem{} Lambeck, Kurt 1980 {\emph{The Earth's Variable Rotation: 
           Geophysical Causes and Consequencies}}, Cambridge University 
           Press, Cambridge, U.K.
\bibitem{} Lambeck, Kurt 1988 {\emph{Geophysical Geodesy}}, Oxford University
           Press, Oxford \& NY
\bibitem{} Landau,  L.D. \& Lifshitz, E.M. 1976 {\emph{Mechanics}},
           Pergamon Press, NY 
\bibitem{} Lazarian, A. \& Draine, B.T., 1997, ApJ, 487, 248  
\bibitem{} Lazarian, A. and Efroimsky, M. 1999, MNRAS, Vol. 303, pp. 673 - 684 
\bibitem{} Marsden, B. \& S. Nakano 1993, IAU Circ. No 5800, 22 May 1993; 
           Marsden, B. \& A. Carusi 1993, IAU Circ. No 5801, 22 May 1993.
\bibitem{} Melosh, H.H. and P. Schenk 1993, Nature, Vol. 365, p. 731
\bibitem{} Nowick, A.S.  and Berry, D.S. 1972 {\emph{Anelastic Relaxation in 
           Crystalline Solids}}, Academic Press, NY 
\bibitem{} Okubo, S. 1982, Geophys.J., Vol. 71, p. 647
\bibitem{} Ostro, S.J., Jurgens, R.F., Rosema, K.D., Whinkler, R.,
           Howard, D., Rose, R., Slade, D.K., Youmans, D.K, Cambell, D.B., 
           Perillat, P, Chandler, J.F., Shapiro, I.I., Hudson, R.S., Palmer, 
           P., and DePater, I., 1993, BAAS, Vol.25, p.1126
\bibitem{} Ostro, S.J., R.S.Hudson, R.F.Jurgens, K.D.Rosema, R.Winkler, 
           D.Howard, R.Rose, M.A.Slade, D.K.Yeomans, J.D.Giorgini, 
           D.B.Campbell, P.Perillat, J.F.Chandler, and I.I.Shapiro, 1995,  
           Science, Vol. 270, p.80-83
\bibitem{} Ostro, S.J., R.S.Hudson, K.D.Rosema, J.D.Giorgini, R.F.Jurgens, 
           D.Yeomans, P.W.Chodas, R.Winkler, R.Rose, D.Choate, R.A.Cormier, 
           D.Kelley, R.Littlefair, L.A.M.Benner, M.L.Thomas, and M.A.Slade
           1999, Icarus, Vol. 137, p. 122-139
\bibitem{} Ostro, S.J., Petr Pravec, Lance A. M. Benner, R. Scott Hudson, 
           Lenka Sarounova, Michael D. Hicks, David L. Rabinowitz, James V. 
           Scotti, David J. Tholen, Marek Wolf, Raymond F. Jurgens, Michael L. 
           Thomas, Jon D. Giorgini, Paul W. Chodas, Donald K. Yeomans, Randy 
           Rose, Robert Frye, Keith D. Rosema, Ron Winkler, and Martin A. 
           Slade, 1999, Science Magazine, Vol. 285, p. 557 - 559
\bibitem{} Peale, S. J., and J. J. Lissauer, 1989, Rotation of Halley's 
           Comet. Icarus, Vol. 79, p. 396-430.
\bibitem{} Purcell, E.M. 1979, Astrophysical Journal, Vol. 231, p.404
\bibitem{} Ryan, M.P.  and Blevins, J.Y.K. 1987 U.S. Geological Survey
           Bulletin, Vol.1764, p.1
\bibitem{} Sekanina Z. 1982, in: {\emph{Comets}}, ed. by L.L.Wilkening,
           University of Arizona Press, Tuscon, pp. 251 - 287
\bibitem{} Scheeres, D.J., S.J.Ostro, R.S.Hudson, S.Suzuki, and E. de Jong 
           1998, Icarus, Vol. 132, p. 53-79
\bibitem{} Spencer, J.W. 1981, J. Geophys. Res., 86, p. 1803.
\bibitem{} Synge, J.L. \& B.A. Griffith 1959 {\emph{Principles of
           Mechanics, Chapter 14}}, McGraw-Hill, NY 
\bibitem{} Tittman, B.R., L. Ahlberg, and J. M. Curnow, 1976, Proc. 7-th Lunar
           Sci. Conf., 3123-3132
\bibitem{} Tschoegl, N.W. 1989 {\emph{The Phenomenological Theory of Linear 
           Viscoelastic Behaviour. An Introduction.}} Springer-Verlag, NY 
\bibitem{} Yoder, C.F. 1982, Icarus, Vol. 49, p. 327

\end{thebibliography}
\end{document}